# THE DEVELOPMENT OF THE JOURNAL ENVIRONMENT OF *LEONARDO*

Alkim Almila Akdag Salah, Virtual Knowledge Studio of the Netherlands Royal Academy of Arts and Sciences, Cruquiusweg 31, 1019 AT Amsterdam, the Netherlands. E-mail: <alelma@ucla.edu> &
Loet Leydesdorff, Amsterdam School of Communications Research (ASCoR), University of Amsterdam, Kloveniersburgwal 48, 1012 CX Amsterdam, the Netherlands. E-mail: <loet@leydesdorff.net >

**Abstract**

We present animations based on the aggregated journal-journal citations of *Leonardo* during the period 1974-2008. *Leonardo* is mainly cited by journals outside the arts domain for cultural reasons, for example, in neuropsychology and physics. Articles in *Leonardo* itself cite a large number of journals, but with a focus on the arts. Animations at this level of aggregation enable us to show the history of the journal from a network perspective.

Scientometric indicators such as citations are increasingly used for the evaluation of performance [1]. In the arts and humanities, however, communication should not be considered primarily as a flow of information [2; cf. 3]. Nonetheless, the *Arts & Humanities Citation Index* (*A&HCI*) of the Institute of Scientific Information (ISI)—currently owned by Thomson Reuters—contains a rich source of statistical information about journals [4]. For example, this data allows us to study a journal's development over time in terms of aggregated cross-journal citation relations.

While this mapping technique has been applied to journals in the domains of the sciences and social sciences, the ISI does not publish an aggregated *Journal Citations Report* (JCR) for the *A&HCI*. In another context, we constructed such an index [5] and focused on the journal *Leonardo* as a prime example of an art journal. We downloaded the bibliographic information for all articles published in *Leonardo* since its inception in 1968 and for all articles citing *Leonardo* during this whole period.

The citations can indicate the impact of *Leonardo* on other (groups of) journals as proxies for fields of scholarly activity, whereas the references provided by the authors in *Leonardo* inform us about what these authors consider the relevant knowledge bases for their publications. Since the retrieved documents did not contain citation information for the period 1970-1973, we limited the analysis to publications since 1974 (5,859 documents). These papers contain 31,147 cited references and were cited 1,680 times. Additionally, we use the 65,285 citations in the citing documents to position *Leonardo* in its citation impact environment. Based on co-citation analysis and bibliographic coupling, respectively and for each consecutive year, we generated animations available online at http://www.leydesdorff.net/journals/leonardo/citing and http://www.leydesdorff.net/journals/leonardo/cited/.

**Figure 1: 53 journals cited by 157 articles in *Leonardo* in 2008; no citation threshold within the set; cosine > 0.0.**

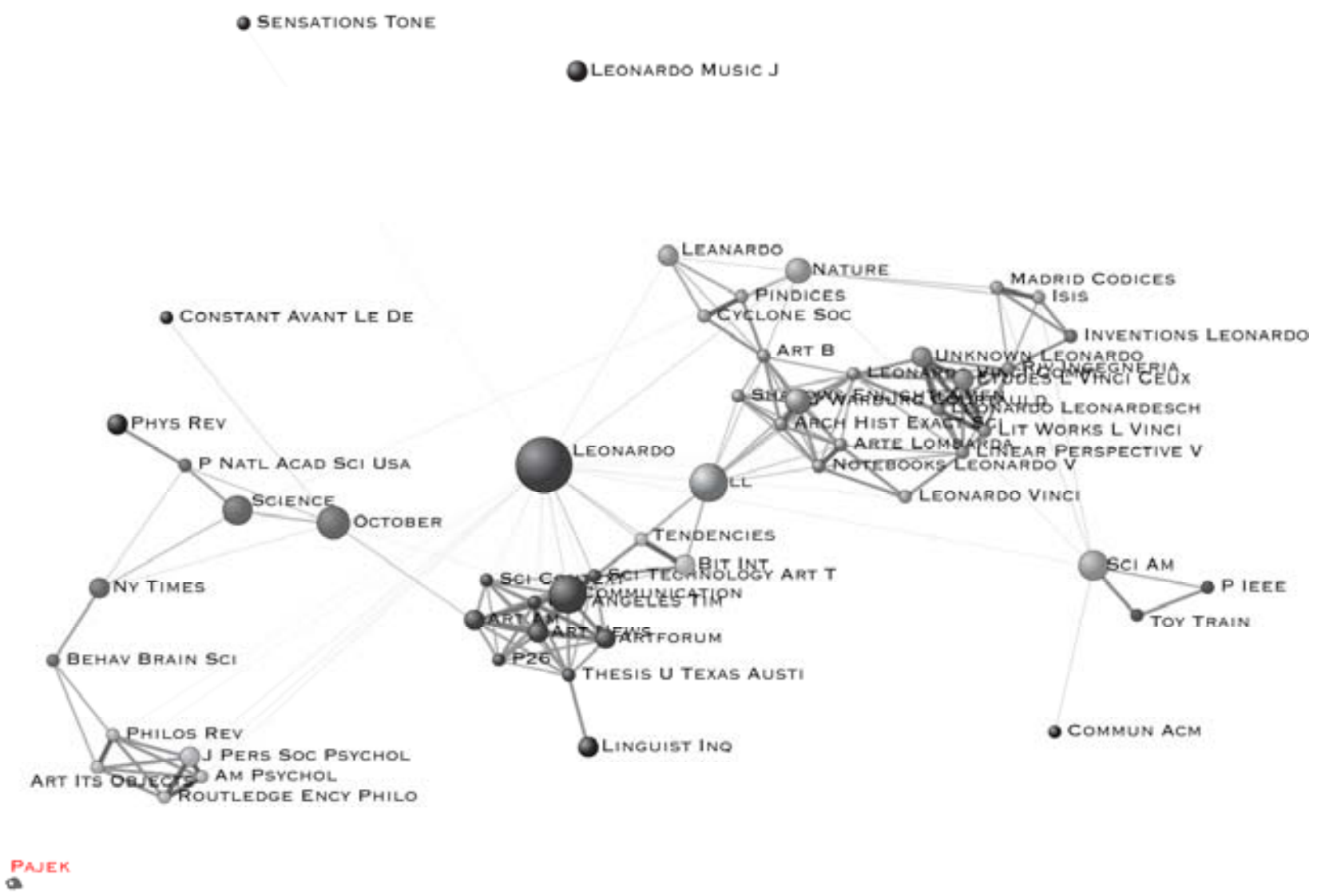

using 107 articles which cited *Leonardo* in 2008.

In the animation showing the journals cited by contributors to *Leonardo*, one can distinguish various art movements, among them Kinetic Art, Holographic Art, Cybernetic Art, Computer Art, and Space Art. In addition to art movements, journals from a wide range of scientific fields are cited—e.g., neurology, cognitive science, psychology, vision and computer graphics—as well as topics such as fractal geometry, applied mathematics, and applied optics. The journal *Health Hazards* is frequently cited, as it contains information about the use of different chemicals and technological appliances in generating artworks.

## Results

These animations locate *Leonardo* as an interdisciplinary journal connected to the sciences, social sciences, and the arts throughout the time span covered. The goal of Frank Malina, the founding Editor of *Leonardo*, was to create a journal enabling artists to follow new developments in the sciences. Editorials in the journal emphasized the interdisciplinary intention and orientation throughout the years in question [6]; our animations show this interdisciplinarity evolving over the years in both the journal's referenced knowledge base ("citing") and its ("cited") impact environment.

The citation patterns are not dense and are therefore volatile from year to year. Figure 1 shows the results of the co-citations of 53 journals cited in 924 references from the 157 articles published in *Leonardo* during 2008. Figure 2 provides the corresponding co-citation map

Art journals form another important group in the animations of the citation patterns of *Leonardo*. This group is more or less equally divided among journals focusing on aesthetics, art theory, and contemporary art news. Upon closer inspection of the animations, one can see that theoretically oriented journals appear more persistently in the animation based on articles citing *Leonardo*, while journals reporting on the latest state of the art market cite *Leonardo* more than they are cited in *Leonardo*'s publications. Core journals of the arts and art history such as *Art News*, *Studio International*, *Art Forum*, *Art Bulletin*, and *Art Journal* are consistently included.

Figure 2 shows an unexpected finding: among the journals which contain documents citing *Leonardo* in 2008, only 24 contribute more than 0.5% to its being cited pattern, and these 24 journals are mainly in the domain of the sciences and the social sciences. Three journal groups are relevant in this citation impact environment: physics, neuroscience, and perception research. These three factors explain 53.8% of the variance. (The nodes are colored in accordance to their highest loadings in the Varimax-rotated three-factor solution.) *Leonardo* itself is positioned at the edge between the latter two specialties.

**Figure 2: Cosine relations among 3,259 references in 107 articles citing *Leonardo* during 2008; only journals which contribute more than 0.5% to the total number of citations; no citation threshold within the set; cosine > 0.0; colors of nodes correspond to the highest factor loadings in a Varimax-rotated three-factor solution.**

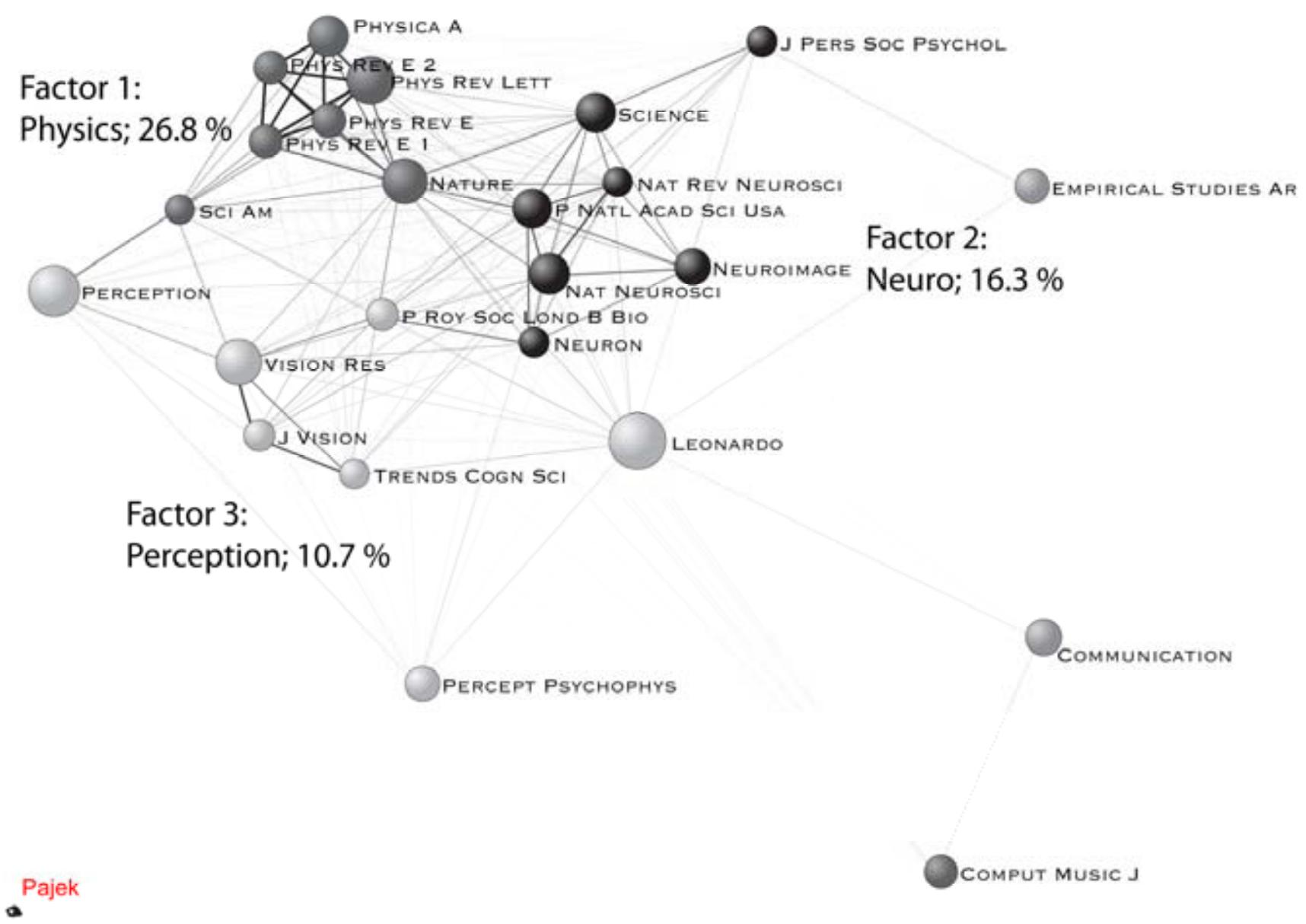

The presence of science journals in the impact environment of *Leonardo* is not stable over the years, but in all years science journals are visible in relatively large clusters. Among the science journals citing *Leonardo* from year to year, *Science*, *Nature*, and *Scientific American* dominate the animation. Since the mid-80s another cluster contains journals with a focus on computer graphics. A third, relatively stable cluster is provided by journals in cognitive science that enter the picture at the beginning of the 1990s, with strong connections to a psychology cluster. Through studies on vision and perception, journals in neuroscience, cognitive science, psychology, and computer graphics are related to this citation environment.

Since the turn of the century, *Leonardo* has increasingly lost citations from the art world in favor of citations from journals in the sciences. In 2008 (Figure 2), science journals dominate its citation impact environment. In earlier years, however, certain core-books by Gombrich [7], Arnheim [8, 9], and Goodman [10] were also cited heavily. These art historians are renowned for their interest in psychology and linguistics, and hence their presence as references in the citation networks strengthened *Leonardo*'s citation relations with journals in these disciplines.

## Generalization and Conclusions

We repeated the analysis of *Leonardo* as a journal for *Art Journal*, a publication of the *College Art Association.* This journal publishes (since 1941) articles related to contemporary art, and in that sense its audience and constituency is akin to that of *Leonardo*. Like *Leonardo*, *Art Journal* is overwhelmingly cited outside the domain of the arts and the humanities. The journals in this larger environment range from physics to advertising research, but most references are to "non-source" journals such as the *NY Times*, *Newsweek*, and the *Washington Post.*

In other words, the impact of journals in the arts is not confined to the *Arts & Humanities* as scholarly discourses in journals, but reaches a much wider audience including the sciences, the social sciences, and the wider public. These journals are cited primarily in the larger environment, perhaps not so much for intellectual as for cultural and instrumental reasons. The predominant rationale of references to these journals is different from that which governs the sciences and the social sciences, where intellectual organization can explain the patterns of citation.

Furthermore, the patterns of citations in the citing and cited dimensions are different for these art journals. Although they draw on a wider environment, it is possible to identify core groups among the journals in the *A&HCI* in terms of how the authors in these journals provide references when constructing their arguments. The citation impact of *Leonardo* on other art journals, however, has decreased over the years.

Given this conclusion, one might indeed be hesitant to assess journals and research covered by the *A&HCI* in terms of scientometric indicators which use field-specific parameters. These journals may occupy positions that are quite different from the specialty structures typical of the sciences and social sciences. Thus, the journals and the constituting articles can be evaluated also in terms of these wider cultural influences. Citation relations are organized not only on socio-cognitive grounds, but also on the basis of cultural patterns.